\documentclass{mem}
\usepackage{natbib}\usepackage{txfonts}\usepackage{balance}
\usepackage{graphicx}
\usepackage[a4paper]{hyperref}
\idline{0}{0}
\begin{document}
\def\teff{$T\rm_{eff }$}
\def\kms{$\mathrm {km s}^{-1}$}

\title{
eROSITA on SRG}

   \subtitle{A X-ray all-sky survey mission}

\author{
N. Cappelluti \,\inst{1} 
\and P. \, Predehl\inst{1} \and H. B\"ohringer\inst{1} \and H. Brunner\inst{1} \and M. Brusa\inst{1} 
\and V. Burwitz\inst{1} \and E.  Churazov\inst{2} \and K. Dennerl\inst{1} \and A. Finoguenov\inst{1} \and M. Freyberg\inst{1}
 \and P. Friedrich\inst{1} \and G. Hasinger \inst{3} \and E. Kenziorra\inst{4} \and I. Kreykenbohm\inst{5}
 \and G. Lamer\inst{6}    \and N. Meidinger\inst{1}  \and M. M\"uhlegger\inst{1} \and M. Pavlinsky\inst{7}    \and J. Robrade\inst{8}
 \and A. Santangelo\inst{4} \and  J. Schmitt\inst{8} \and A. Schwope\inst{6} \and M. Steinmitz\inst{6} \and L. Str\"uder\inst{1} \and R. Sunyaev\inst{1}
 \and C. Tenzer\inst{4} }

  \offprints{N. Cappelluti}

\institute{
Max-Planck-Institut f\"ur extraterrestrische Physik, D-85741 Garching, Germany
\and
Max-Planck-Institut f\"ur Astrophysik, D-85741 Garching, Germany
\and
Max-Planck-Institut f\"ur Plasma Physik, D-85741 Garching, Germany
\and
Institut f\"ur Astronomie und Astrophysik, Abteilung Astronomie, Universit\"at T\"ubingen, Sand 1, 72076 T\"ubingen, Germany
\and
 Universit"at Erlangen/N\"urnberg, Dr.-Remeis-Sternwarte Bamberg, Sternwartstrasse 7, D- Bamberg
 \and
 Astrophysikalisches Institut Potsdam, An der Sternwarte 16, D-14482 Potsdam, Germany
 \and
 Space Research Institute Moscow, Russian Federation
 \and
 Universit\"at Hamburg, Hamburger Sternwarte, Gojenbergsweg 112, 21029 Hamburg, Germany
\email{cap@mpe.mpg.de}
}

\authorrunning{Cappelluti, N. et al.}

\titlerunning{eROSITA on SRG}

\abstract{
eROSITA (extended ROentgen Survey with an Imaging Telescope Array) is the core instrument on the Russian 
Spektrum-Roentgen-Gamma (SRG) mission which is scheduled for launch in late 2012. eROSITA is fully approved and funded 
by the German Space Agency DLR and the Max-Planck-Society.

The design driving science is the detection of 50 - 100 thousands Clusters of Galaxies up to redshift $z\sim$1.3 in 
order to study the large scale structure in the Universe and test cosmological models, especially  Dark Energy. 
This will be accomplished by an all-sky survey lasting for four years plus a phase of pointed observations. 
eROSITA consists of seven Wolter-I telescope modules, each equipped with 54 Wolter-I shells
having an outer diameter of 360 mm. This would provide and effective area at 1.5 keV of $\sim$ 1500 cm$^{2}$ and
an on axis PSF HEW of 15$^{\prime\prime}$ which would provide an effective angular resolution of
25$^{\prime\prime}$-30$^{\prime\prime}$. In the focus of each mirror module, 
a fast frame-store pn-CCD will provide a field of view of 1$^{\circ}$ 
in diameter for an active FOV of $\sim$0.83 deg$^{2}$. 
At the time of writing the instrument development is currently in phase C/D.
}
\maketitle{}

\section{Mission overview}

The Russian Spectrum-Roentgen-Gamma (SRG) satellite  will fly on  a medium class platform 
(''Navigator'', Lavochkin Association, Russia). 
The launch will be in 2012 using a Soyuz-2 rocket from Bajkonur into an orbit around L2. 
The payload consists of the X-ray instruments eROSITA (extended ROentgen Survey with an Imaging Telescope Array) 
and ART-XC (Astronomical Roentgen Telescope -- X-ray Concentrator).

The seven eROSITA telescopes are based on the existing design launched on the ABRIXAS mission plus 
an advanced version of the pnCCD camera successfully flying on XMM-Newton. In order to optimize eROSITA 
for the Dark Energy studies, the effective area is increased by a factor of five, 
the angular resolution is improved by a factor of two, and the field of view is also increased by a factor of two
with respect to ABRIXAS. 
Such a design has been drawn to match the outcome of the most recent calls  for 
ideas on Dark Energy observations (like e.g. by NASA, DOE, ESA, ESO and others). 

Similarly to eROSITA ART-XC contains 7 telescopes working in the energy range between 6 and 30 keV. 
The telescopes are conical approximations of the Wolter-I geometry with CdTe detectors in their focal 
planes (Pavlinsky et al., 2010, this Volume)

\section{Design Driving Science}
\subsection{Dark Energy}
\begin{figure*}[!t]
  \includegraphics[scale=.35]{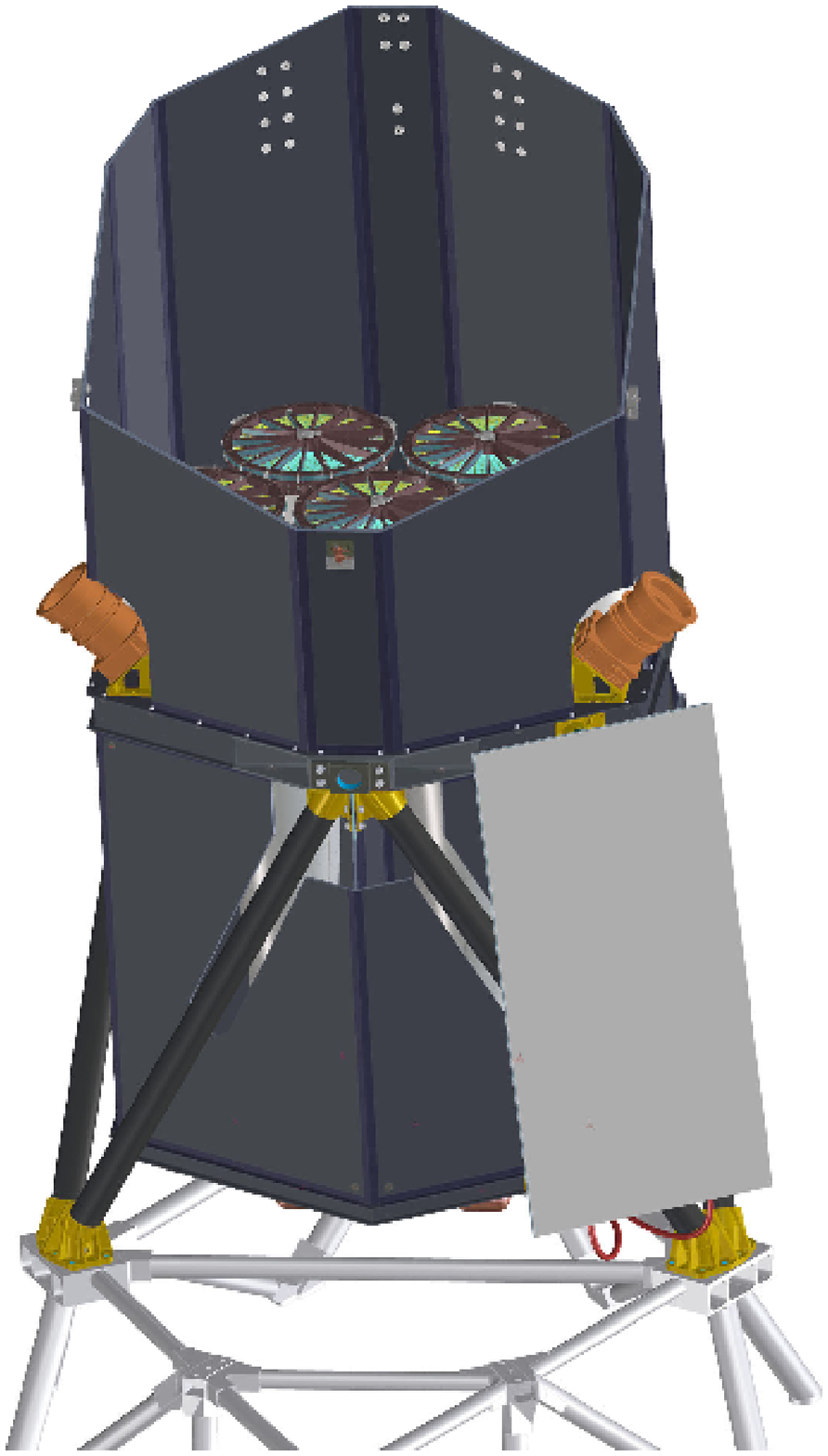}
 \includegraphics[scale=0.35]{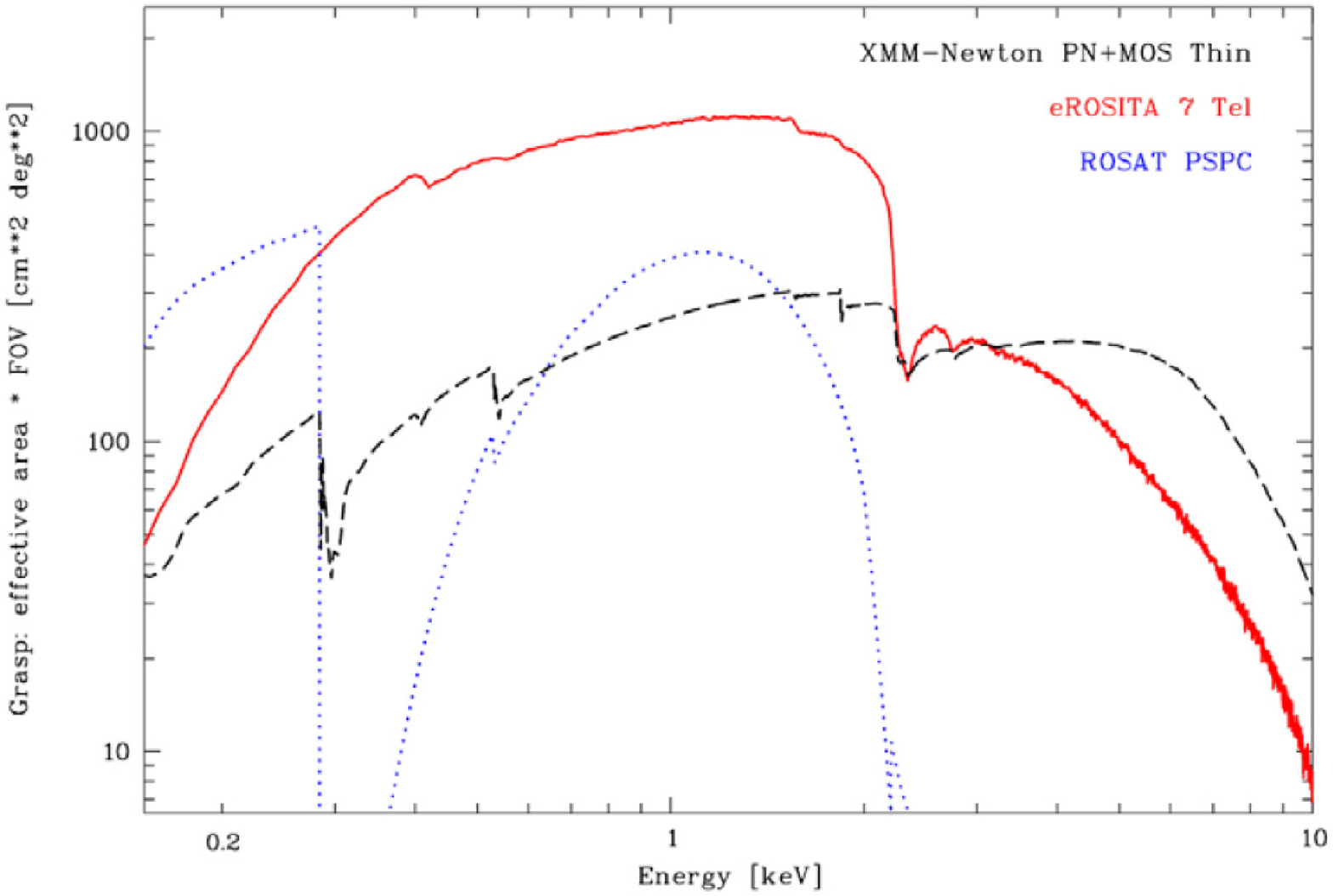}
  \caption{$Left~Panel:$ The eROSITA telescope on board SRG. $Right~Panel:$
The grasp of eROSITA compared with ROSAT-PSPC and XMM-Newton.}
  \label{grasp}
\end{figure*}
One way to test cosmological models and to assess the origin, geometry, and dynamics of 
the Universe is through the study of the large-scale structures
Indeed galaxy clusters are strongly correlated and thus they are good tracers
 of the large-scale structure on very large scales
by sampling the most massive  congregates of matter. 
The galaxy cluster population provides information on the cosmological 
parameters in several complementary ways:
\begin{figure*}[!t]
  \includegraphics[width=\textwidth]{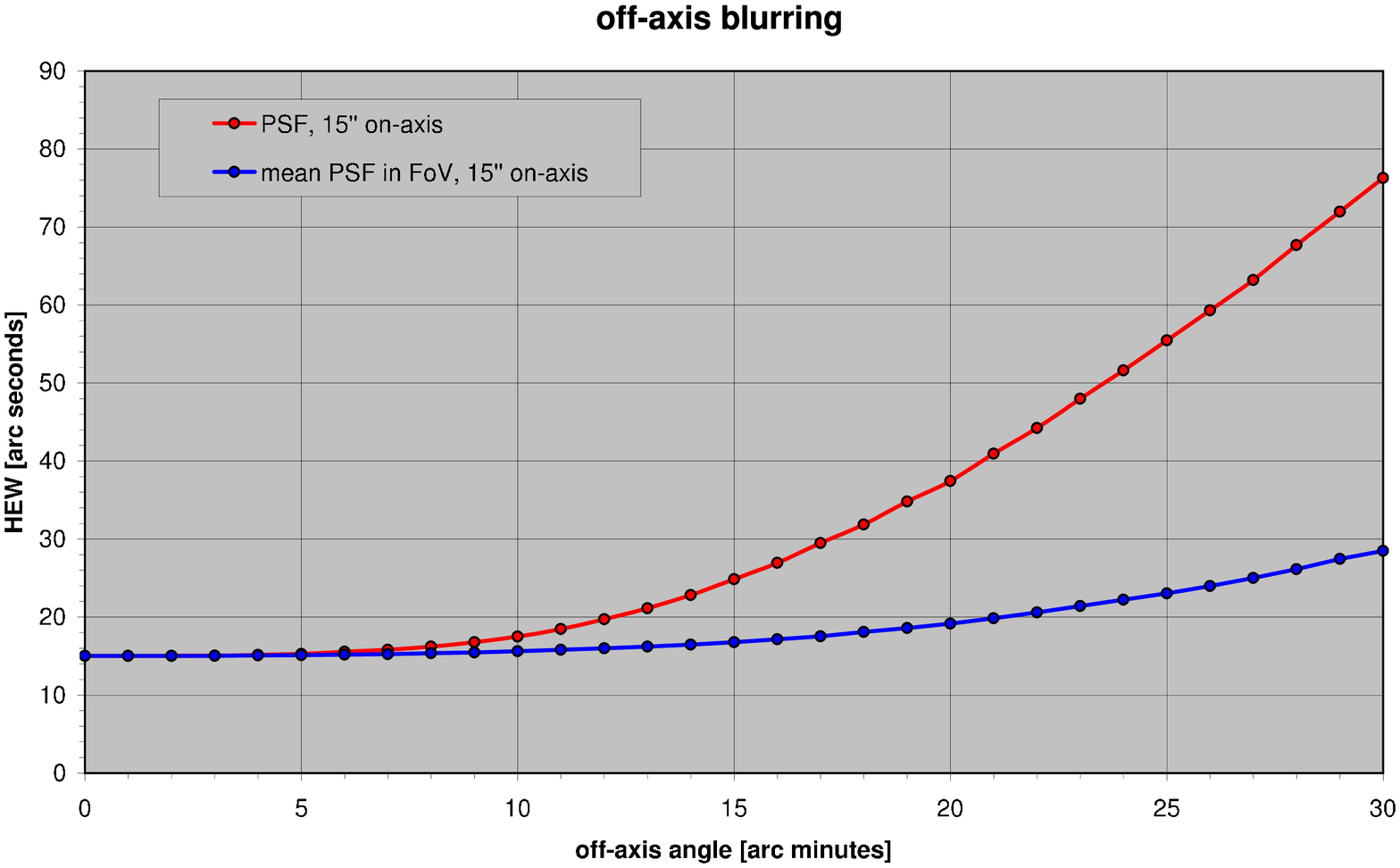}
  \caption{$Red~Curve$: the PSF HEW as function of the offaxis angle in pointing mode. $Blue~Curve$: the
PSF HEW within an encircled off-axis angle during a scan. Basically the value at 30$^{\prime}$, 
is the cumulative PSF in the survey. }
  \label{grasp-fov}
\end{figure*}
\begin{enumerate}
\item
  The cluster mass function in the local Universe mainly depends on the matter density
  $\Omega_m$ and the amplitude of the primordial power spectrum  $\sigma_8$.
\item
  The evolution of the mass function f(M,z) is 
directly determined by the growth of structure in the Universe and therefore gives 
at the same time sensitive constraints on Dark Matter and Dark Energy.
\item
  The amplitude and shape of the cluster power spectrum , P(k) and its growth with time,  
depend sensitively on Dark Matter and Dark Energy.
\item
  Baryonic wiggles due to the acoustic oscillations at
 the time of recombination are still imprinted on the large 
scale distribution of clusters (i.e. in their P(k) and the Autocorrelation function) 
and thus can give tight constraints on the curvature of space at different epochs.
\end{enumerate}
The constraints provided by the different cosmological tests with clusters
are complementary in such a way, that degeneracies in the parameter constraints 
in any of the tests can be broken by combinations. 
The simultaneous constraint of  $\Omega_m$ and  $\sigma_8$  by combining method 1 and 3 above
 is one such example \cite{Schuecker}. In addition the combination
 of several tests provides important consistency checks as explained below. 
In addition to the above applications, galaxy clusters have been used as 
cosmological standard candles to probe absolute distances,
 analogous to the cosmological tests with supernovae type Ia.
 The assumption that the cluster baryon fraction is constant with 
time combined with observations of this quantity provides constraints 
on Dark Matter and Dark Energy,   e.g. \cite{Allen}).
In a very similar way, combined X-ray and Sunyaev-Zeldovich-measurements 
provide a mean for absolute distance measurements and constraints of the geometry 
of the Universe, e.g. \cite{Molnar}.\\
Large, well defined and statistically complete samples of galaxy clusters 
(which are dynamically well evolved and for which masses are approximately known) are
obvious prerequisites for such studies. 
A substantial progress in the field requires samples of tens to hundreds 
of thousands of clusters. Surveys at several wavelengths are used or planned to be used to achieve this goal. 
More in general, in X-ray surveys, galaxy clusters are detected by the radiation of the hot intracluster medium and 
X-ray observations are still the most efficient tools to  select  clean cluster samples. In addition X-ray observations 
provide accurate estimates of the cluster physical parameters. Indeed
the X-ray luminosity is tightly correlated to the gravitational mass, temperature and core radius \cite{Reiprich}.
Therefore most cosmological studies involving galaxy clusters 
are based on X-ray surveys (e.g. \cite{Henry00}, \cite{Henry04},
 \cite{boeringer}, \cite{Vikhlinin}. 
\section{Instrument}
The mirror replication technique was developed for XMM-Newton and has then been 
applied to the small satellite mission ABRIXAS, which had scaled the XMM-Newton telescopes 
down by a factor of about 4. The ABRIXAS optical design and
 manufacturing process are adopted for eROSITA partially because 
the inner 27 mirror shells and therefore the focal length are kept the same. 
The mirror system consists of 7 mirror modules with 54 mirror shells each and a 
X-ray baffle in front of each module. Unlike on ABRIXAS, the seven optical 
axes are co-aligned. Compared to a large single mirror system, 
the advantages of a multiple mirror system are: shorter focal length 
(reduced instrumental background) and reduced pileup when observing bright sources.
 This configuration allows a more compact telescope and multiple but identical 
cameras which automatically provides a 7-fold redundancy. 
The capabilities of the X-ray mirror system are described by effective area, 
vignetting function, and PSF.  The production of the flight mirrors has already started.
The eROSITA-CCD \cite{Meidinger} have 384 $\times$ 384 pixels or an image area of 28.8 mm  $\times$ 28.8 mm, 
respectively, for a field of view of 1.03$^{\circ}$ diameter. The 384 channels are read out in parallel. 
The nominal integration time for eROSITA will be 50 msec. The integrated image can be shifted into the 
frame store area by less than 100 msec before it is read out within about 5 msec. 
CCD together with the two CAMEX and the (passive) front-end electronics are integrated on a ceramic
 printed circuit board (CCD-module) and is connected to the "outer world" by a flexlead. 
The flight-CCDs have already  been fabricated. For operation the CCDs have to be cooled down
 to -80 $^{\circ}$C by means of passive elements (heatpipes and radiator). 
Fluorescence X-ray radiation generated by cosmic particles is minimized by a 
graded shield consisting of aluminum and boron carbide. For calibration purposes, 
each camera housing contains a radioactive Fe$^{55}$ source and an aluminum target providing 
two spectral lines at 5.9 keV (Mn-K$_{\alpha}$ ) and 1.5 keV (Al-K$_{\alpha}$). 
The mechanism ("Filter Wheel") for moving the calibration source into and out
 of the field of view is designed and qualified. Also the telescope structure is also qualified.
 The optical bench connects the mirror system and the baffles on one side with 
the focal plane instrumentation on the other side. Additionally it forms the mechanical
 interface to the S/C bus.  The flight model manufacturing is ongoing. 
The dimensions of the telescope structure is of the order 
 1.9 m diameter x 3.2 m height. The total weight of eROSITA is  735 kg \cite{Predehl}.
The instrument design is shown in Figure \ref{grasp}.


\section{Sensitivity}

Figure \ref{grasp} shows the grasp of eROSITA, i.e. the product of effective area and solid angle of the field of view. 
The effective area of eROSITA is about twice that of one XMM-Newton telescope in the energy band below 2keV, 
whereas it is three times less at higher energies. This is a consequence of the small f-ratio 
(focal length vs. aperture) of the eROSITA mirrors. An advantage of the short focal length 
is also larger field of view.  The eROSITA angular resolution is 
15 arcsec on-axis. Due to the unavoidable off-axis degradation of a Wolter-I telescope,
 the angular resolution averaged over the field of view is of the order of  
28$^{\prime\prime}$ (Fig. 2). We will scan the entire sky for four years (ROSAT 1/2 year). 
Therefore the eROSITA sensitivity during this all-sky survey will be approximately 30 times ROSAT.
With the current scanning strategy, we expect an average exposure of $\sim$3 ks in the all-sky survey,
with two deep fields at the ecliptic poles with an exposure of the order of 20-40 ks, depending on the actual
mission strategy.

We have performed simulations of the radiation environment in L2 and determined,
by including the cosmic components, a background intensity of 5.63 cts s$^{-1}$ deg$^{-2}$ and
3.15 cts s$^{-1}$ deg$^{-2}$ in the 0.5--2 keV and 2--10 keV energy bands, respectively.

The 0.5-2 keV flux limit for clusters will be, on average, of the order of 3$\times$10$^{-14}$   erg cm$^{-2}$ s$^{-1}$
and 5$\times$10$^{-15}$   erg cm$^{-2}$ s$^{-1}$ in the all-sky survey and in the ecliptic poles, respectively.
In Figure 3 we plot the eROSITA 5$\sigma$ point source flux limit of the survey in the 0.5--2 keV and 2--10 keV energy bands 
as  function of the exposure time. In the all-sky survey the typical flux limit will be $\sim$10$^{-14}$
 erg cm$^{-2}$ s$^{-1}$ and $\sim$3$\times$10$^{-13}$  erg cm$^{-2}$ s$^{-1}$ in the 0.5-2 keV and 2--10 keV 
energy band, respectively. At the poles we expect to reach flux limits of the order of 
 $\sim$2$\times$10$^{-15}$
 erg cm$^{-2}$ s$^{-1}$ and $\sim$3$\times$10$^{-14}$  erg cm$^{-2}$ s$^{-1}$ in the 0.5-2 keV and 2--10 keV 
energy band, respectively. Note that the observation will be photon limited up to exposures of $\sim$20 ks. 
In the 0.5--2 keV band, the confusion limits of 1 source every 10 beams will be reached in about 20-30 ks.
 At this fluxes the X-ray sky is dominated 
by clusters and AGN, which can be separated with an angular resolution of 25$^{\prime\prime}$--30$^{\prime\prime}$. 
The logN-logS of  clusters is well known to the proposed depth (\cite{Gioia}, \cite{Rosati}, \cite{fin}). 
The proposed survey will identify 50,000~100,000 clusters depending on the capabilities in 
disentangle moderately-low extended sources from AGN. Concerning the  number of 
AGN we can use the  logN-logS measurement in moderately wide field surveys,
like XMM-COSMOS \cite{cap07}, \cite{cap09},
 to predict the detection 3-10$\times$10$^{6}$ sources,
 up to z$\sim$7-8, depending on the detection threshold. A simulation of a 3 ks eROSITA observation 
of a  typical extragalactic field is shown in Figure 3. 
Multi-band optical surveys to provide
the required photometric and spectroscopic redshifts are already in the planning stages, and will be contemporaneous
 with or precede  our survey. The cluster population will essentially cover the redshift range z = 0 - 1.3 
and will reveal all evolved galaxy clusters with masses above $3.5 \times 10^{14} h^{-1} M_{\odot}$ up to redshifts of 2. 
Above this mass threshold the tight correlations between X-ray observables and mass allow direct interpretation 
of the data.This sample size is necessary for example to precisely characterize the cluster mass function and 
power spectrum in at least ten redshift bins, to follow the growth of structure with time.
 \begin{figure*}[!t]
  \includegraphics[scale=0.27]{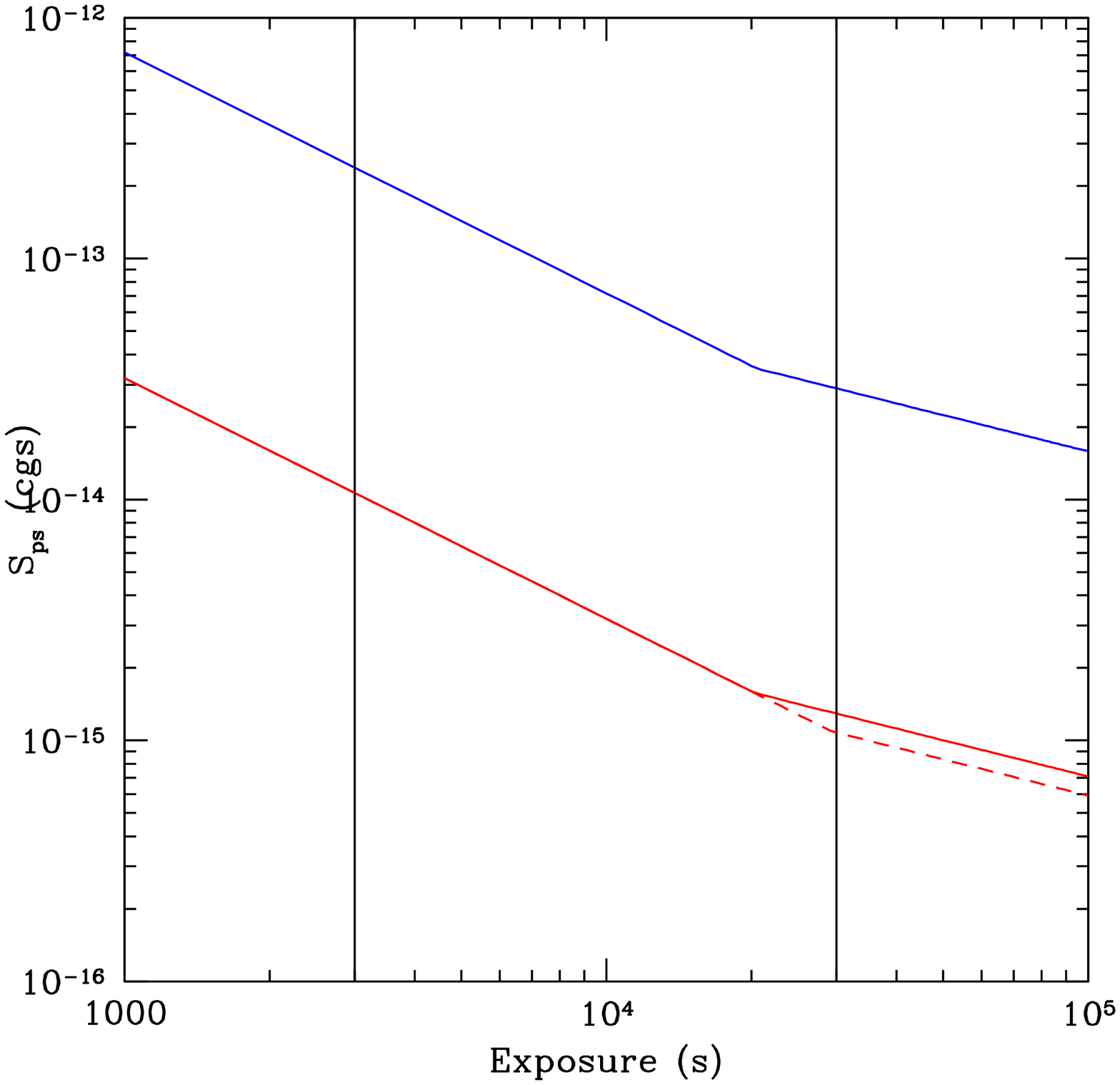}
 \includegraphics[scale=0.45,]{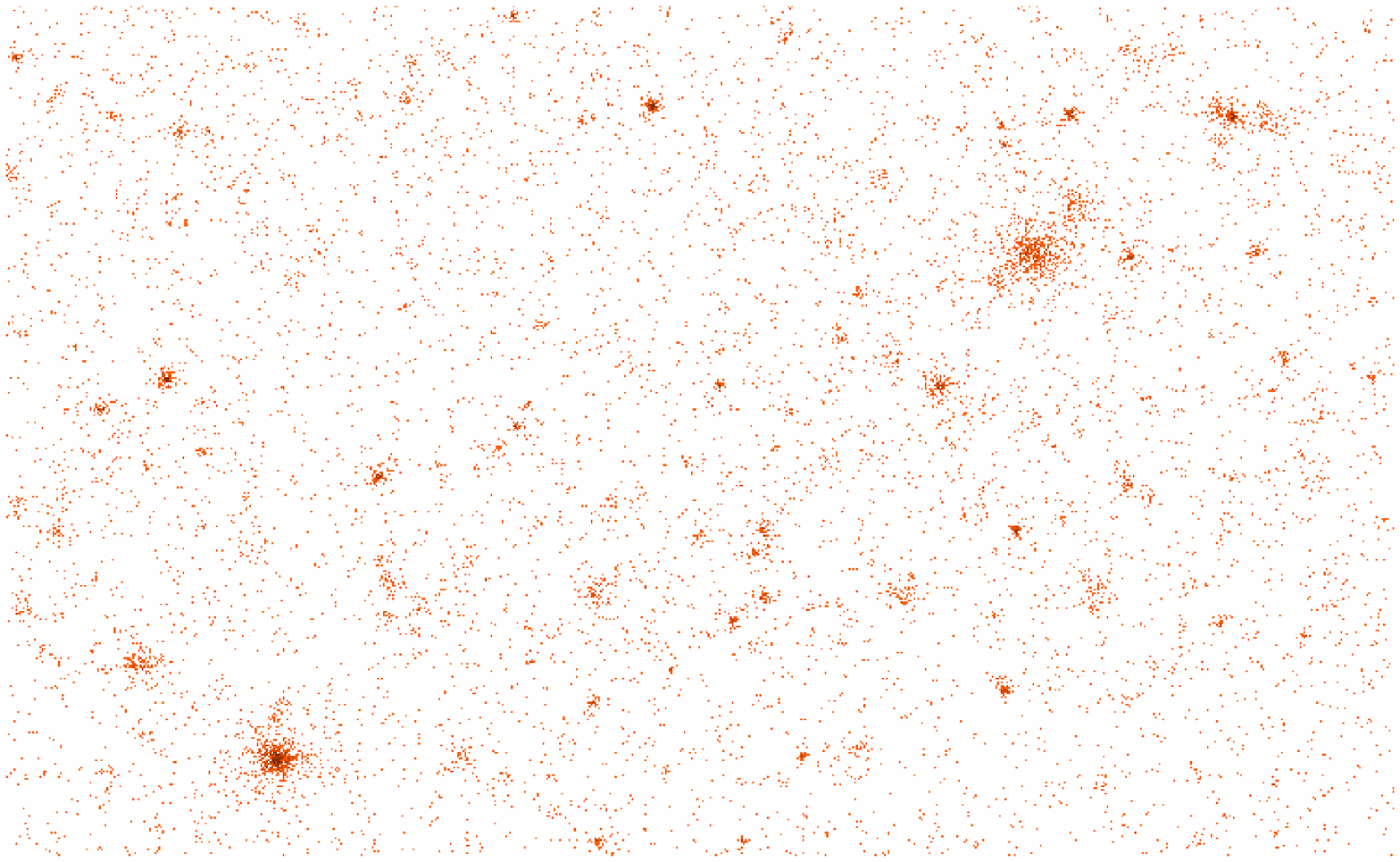}  
  \caption{$Left~Panel:$ the 5$\sigma$ point-source sensitivity vs. exposure in the 0.5--2 keV ($red$)
and 2--10 keV ($blue$) energy bands by assuming a PSF HEW of 30$^{\prime\prime}$ and 40$^{\prime\prime}$, respectively.
  The dashed lines is the sensitivity achieved with an average PSF-HEW of 25$^{\prime\prime}$. $Right~Panel:$ A 
simulation of a 3ks observation of a 1$^{\circ}\times$1.6$^{\circ}$ of eROSITA in survey mode. The simulations
includes cosmis+particle background, randomly distributed AGN and clusters extracted from Hydrodinamical simulations.}
  \label{sensitivity}
\end{figure*}



\end{document}